\documentclass[12pt,preprint]{aastex}

\begin{document}

\newcommand{\beq}{\begin{equation}}
\newcommand{\beqa}{\begin{eqnarray*}}
\newcommand{\beqan}{\begin{eqnarray}}
\newcommand{\greq}{\begin{equation}\left\{ \begin{array}{l}}
\newcommand{\eeq}{\end{equation}} 
\newcommand{\eeqa}{\end{eqnarray*}}
\newcommand{\eeqan}{\end{eqnarray}}
\newcommand{\hsp}{ \hspace{.5cm} }
\newcommand{\noi}{ \noindent }
\newcommand{\ssi}{ \Longleftrightarrow}
\newcommand{\lp}{ \left(}
\newcommand{\rp}{ \right)}
\newcommand{\lc}{ \left[}
\newcommand{\rc}{ \right]}
\newcommand{\dzeta}{\zeta}
\newcommand{\khi}{\chi}
\newcommand{\YL}{ Y^m_l }
\newcommand{\PL}{ P^m_l }
\newcommand{\eps}{\varepsilon}
\newcommand{\na}{ \vec{\nabla} }
\newcommand{\intsur}{ \int_{(S)}\! }
\newcommand{\intvol}{ \int_{(V)}\! }
\newcommand{\cth}{ \cos\theta }
\newcommand{\sth}{ \sin\theta }
\newcommand{\dOm}{d\Omega}
\newcommand{\vvphi}{\vec{v}_{\phi}}
\newcommand{\demi}{\frac{1}{2}}
\newcommand{\omb}{\overline{\omega}}
\newcommand{\rhobar}{\overline{\rho}}
\newcommand{\ephi}{\vec{e}_\phi}
\newcommand{\ddr}[1]{\frac{{\rm d}  #1}{{\rm d} r}}
\newcommand{\dr}[1]{\frac{\partial  #1}{\partial r}}
\newcommand{\dt}[1]{\frac{\partial  #1}{\partial t}}
\newcommand{\dnt}[1]{\frac{{\rm d}  #1}{{\rm d}t}}
\newcommand{\dnx}[1]{\frac{{\rm d}  #1}{{\rm d}x}}
\newcommand{\dny}[1]{\frac{{\rm d}  #1}{{\rm d}y}}
\newcommand{\dnz}[1]{\frac{{\rm d}  #1}{{\rm d}z}}
\newcommand{\Dt}[1]{\frac{D #1}{D t}}
\newcommand{\dx}[1]{\frac{\partial  #1}{\partial x}}
\newcommand{\dy}[1]{\frac{\partial  #1}{\partial y}}
\newcommand{\dz}[1]{\frac{\partial  #1}{\partial z}}
\newcommand{\dtheta}[1]{\frac{\partial  #1}{\partial \theta}}
\newcommand{\drtheta}[1]{\frac{1}{r}\frac{\partial  #1}{\partial\theta}}
\newcommand{\ddrr}[1]{\frac{\partial^2  #1}{\partial r^2}}
\newcommand{\ddtt}[1]{\frac{\partial^2  #1}{\partial t^2}}
\newcommand{\drr}{\frac{\partial}{\partial r}}
\newcommand{\dxx}{\frac{\partial}{\partial x}}
\newcommand{\dtt}{\frac{\partial}{\partial t}}
\newcommand{\ksi}{\xi}

\title{Layer Formation in Semiconvection}

\author{Joseph A. Biello}

\affil{Department of Astronomy and Astrophysics, The University of Chicago,
	Chicago, IL, 60637, USA 
\footnote{current address: Department of Mathematical Sciences,
Rensselear Polytechnic Institute, Troy, NY 12180, USA}
}
\email{biello@oddjob.uchicago.edu}
\shorttitle{Layer Formation in Semiconvection}
\shortauthors{J. A. Biello}

\begin{abstract}
Layer formation in a thermally destabilized fluid with stable 
density gradient has been observed in laboratory experiments 
and has been proposed as a mechanism for mixing molecular
weight in late stages of stellar evolution
in regions which are unstable to semiconvection.
It is not yet known whether such layers can exist in a very
low viscosity fluid: this work undertakes to address that question.
Layering is simulated numerically both at high Prandtl number
(relevant to the laboratory) in order to describe the onset of
layering intability,
and the astrophysically important case of low  Prandtl number.
It is argued that the critical stability parameter for
interfaces between layers, the Richardson number, increases with
decreasing Prandtl number. 
Throughout the simulations the fluid has a tendency to form
large scale flows in the first convecting layer, but only 
at low Prandtl number do such structures
have  dramatic consequences for layering.  These flows 
are shown to drive large interfacial waves whose breaking
contributes to significant mixing across the interface. 
An effective diffusion
coefficient is determined from the simulation
and is shown to be much greater than the predictions of
both an enhanced diffusion model and one which specifically
incorporates wave breaking.  The results further suggest that 
molecular weight gradient interfaces are ineffective barriers
to mixing even when specified as initial conditions, such
as would arise when a compositional gradient is redistributed
by another mechanism than buoyancy, such as rotation
or internal waves.

\end{abstract}




\section{Introduction}

Though semiconvection may be an important mechanism for mixing of
chemicals in late stages of stellar evolution, there have been
few numerical simulations of semiconvective scenarios.  
In part, this is due to the fact that the numerical resolution
required for a simulation has, until recently, made computation
prohibitive.  Coupled to that, the fact that  the existing mixing
length theories (such as  \citet{stevenson}, 
\citet{langer_etal} and \citet{spruit}) 
which both draw on theoretical
and experimental arguments, seemed adequate when compared to observations.

However, recent observations of SN1987A have been at odds with models of
late stages of stellar evolution and one main suspect
are the theories of elemental mixing used in those models 
(\citet{maeder}). 
As with all turbulence models, the more sophisticated
semiconvective models 
(\citet{canuto_99}, \citet{grossman}, \citet{grossman_etal})
which arose in the wake of these new data require
closure assumptions.  Implicit in these are assumptions about the character
of the flow which have yet to be tested against numerical experiments
of semiconvection.  
Numerical simulations are now sophisticated enough that 
we can hope to distinguish between the assumptions made in the 
mixing length theories and settle the question of the heat transport,
chemical mixing and velocity field associated with semiconvective zones.
In this work, I shall address one particular question related to all 
three of these topics: does layered convection 
develop when a compositionally stratified fluid, with very low viscosity, 
is subjected to a vertical heat flux?

\subsection{Semiconvection and Double Diffusion}

\citet{kato_1} performed a 
local analysis of the convective instability of
 gases with a varying compositional gradient subjected to 
an upward heat flux and showed three criteria for instability.
If the molecular weight increases upward, $d \mu /dz > 0$, the fluid is
unstable to the Rayleigh-Taylor instability.  If 
\beq
\nabla_{rad} > 
\nabla_{Led} = \nabla_{ad} + \frac{\beta}{4-3\beta} \nabla_{\mu}
\label{ledoux_crit}
\eeq
then the fluid is exponentially
unstable to classical convection.  
Here, $ \nabla_{ad} $ is the adiabatic temperature gradient,
$\nabla_{ad}= (\partial \ln T/\partial \ln p)_S$
and $\nabla_{rad}$  is the corresponding temperature derivative
due to radiative diffusion.
$\nabla_{\mu}= \partial \ln \mu/\partial \ln p$, $\mu$ is the
molecular weight and $\beta$ is the ratio of gas pressure to the total
pressure. 
This criterion was discovered by \citet{ledoux} and
bears his name.  Finally, Kato demonstrated that when 
\beq
\nabla_{ad}< \nabla_{rad} < \nabla_{Led} 
\label{schwarz_crit}
\eeq
convection sets in through overstable oscillations.  
The physical explanation of this instability
requires considering  blobs of fluid 
rising in a hydrostatically stratified medium. If the medium
is stable to the Ledoux criterion, such blobs will fall back to their
initial positions.  If, further, we allow heat to diffuse from 
the blobs (by far the fastest diffusing process in astrophysical
contexts) then the returning blobs will have lost a small amount of heat
on their round trip and will overshoot their initial positions.  
This mechanism relies on the existence of thermal
diffusion and the growth rate of overstable modes is
in fact proportional to the thermal diffusivity.   
Thus, overstable convection grows much more slowly than 
classical thermal convection and it would be reasonable to
expect that the convection, itself, is less vigorous than
its classical counterpart.  

The weakness of linear oscillatory convection may be 
a completely irrelevant fact since the bifurcation
that leads to overstability is subcritical.  That
is to say, the fluid can go unstable to small, but finite
amplitude perturbations, at values of the temperature gradient
far below that required for overstability.
This is a weakly non-linear effect and is due to 
the fact that a small rearrangement in the molecular 
weight (due to breaking gravity waves,
shear instability, etc) can drive the
fluid to classical thermal convection, since the 
stabilizing compositional gradient
may be locally destroyed.  

Weakly non-linear analysis was performed by
\citet{veronis_68} for an incompressible, Boussinesq
fluid with a stabilizing  salt gradient and destabilizing temperature
gradient.  Though superficially appropriate for only the oceanographic
situation for which it was derived, it was pointed
out by E.A. Spiegel (see \cite{kato_1}) that the 
linear stability equations
of semiconvection are equivalent to the incompressible equations with
the salt playing the role the molecular weight.
In the oceanographic context such overstability
is referred to as double diffusion.

The effect of the redistribution of the stabilizing component
is most dramatically seen in the now famous experiment by 
\citet{turner_stom} and reproduced by
\citet{turner_68},
\citet{huplin} 
and \citet{fern_87} and \citet{fern_89}. 
In the prototype and all of its successors, 
a tank of water with a stable salinity concentration gradient is
heated from below.  At first the water becomes 
unstable to rapidly rising convective plumes but, since
they are rising into a region of less salt concentration, the
plumes are restricted to a thin region at the bottom of the tank.  The salt is
rapidly mixed thus there is no gradient 
in this region and the fluid motion becomes turbulent.  Subsequently,
the layer proceeds to slowly entrain from the nonconvecting
fluid above it while transporting heat to that
region and  at all times increasing in temperature.
After a critical layer thickness is reached, the 
boundary layer separating the convecting fluid from the
quiescent fluid above is seen to undergo oscillatory instability.  This, 
in turn, redistributes the salt in the second convecting layer
which is now separated from the first by a buoyancy jump.  The convection 
becomes turbulent in this second layer and the process continues in
subsequent layers.  
At intermediate times, the tank has formed a vertical stack of convecting
layers with well mixed temperature and salt within each.  
Though mixing between layers is limited by the saline density jump across each,
the lower layers are able to slowly erode this density jump 
over much longer time scales by the breaking
of internal waves on the interfaces between the layers. 

\citet{fern_87} \& \citet{fern_89}, provide the most accurate
quantitative data for the water-salt experiment.
In particular, he confirms the existing theory for the development
of the first layer and gives a criterion for the stability of a layer.
Defining $l$ as the integral length scale of the turbulence, the scale
at which the turbulence is forced, $u_*$ as the root mean square turbulent
velocity in the convecting layer, $\rho_*$,  the
average density of the convecting layer and $\Delta b = g (\Delta \rho) $ 
as the buoyancy contrast
between the convecting layer and the fluid above, then
\beq
Ri = \frac{ l \,\Delta b}{\rho_* u_* ^2}
\label{eq:richardson}
\eeq
is the Richardson number of the flow.  This parameter generally 
arises in the study of a stable buoyancy gradient in the presence of a 
horizontal shear flow.  In the layered convection case it amounts to
the ratio of the potential energy difference over the scale of the
largest eddies in the turbulent flow to the kinetic energy of the flow.
Fernando showed that, in the case of water-salt experiment, 
there are three regimes of interest delimited by the Richardson number.

At low Richardson numbers, the turbulent eddies are able to penetrate
deeply into the static fluid and the interface migrates upward.  At intermediate
Richardson numbers, mixing across the stable density interface
is primarily due to the breaking of interfacial (internal) waves and the
scouring of this surface by a weaker flow that can be maintained in the upper
fluid.  At higher Richardson number, the effect of interfacial waves and penetration
becomes negligible as convective plumes flatten out when they hit the density interface.  
Transport across the layers then takes place purely by diffusion, though enhanced by
the steep gradient at the interface.  

Only when the Richardson number, based on the uppermost layer and the static fluid,
is of intermediate or high values does there develop an oscillatory instability 
in the boundary between those two regions: another layer forms.  

Let us now return to astrophysical semiconvection, contrasting 
the semiconvective theories of \citet{spruit} and \citet{stevenson}.  
Using the intuition of the laboratory experiments, Spruit constructed 
a mixing length theory of layered convection separated by thin molecular
weight gradients and thicker thermal gradients.  Transport across the layers
is solely due to diffusion through the thin thermal 
and helium transition regions at the interfaces of the layers.  
In fact, this theory is similar to the double diffusive theory constructed for
confined geometries by \citet{knob_merry}.
However, confined geometries cannot
support surface waves like the internal waves in layered convection and
these could contribute to even greater cross interface transport.

\citet{stevenson} described 
a theory of semiconvection in which growing overstable modes 
resonantly feed energy to
smaller scales and eventually break, 
whereupon the molecular weight gradient is redistributed.  
By making reference to the salt-water experiments and
energetic arguments for double diffusive convection, he compellingly argued
that, should layers form as a result of this wave breaking, they would be unstable
if $Pr < \tau^{1/2}$.  The Prandtl number, defined as $Pr = \nu / \kappa_T$, 
and the Lewis number, 
$\tau = \kappa_{He} / \kappa_T$, are ratios of the  kinematic viscosity and
molecular diffusivity of the second species to the thermal diffusivity, respectively.
(The two species are helium in a hydrogen convection zone.)
In semiconvection, the thermal diffusivity is dominated by radiation whereas both
viscosity and helium diffusion are molecular processes and, as such, $1 \gg Pr  > \tau$.
In contrast to Spruit, Stevenson's theory would lead to the conclusion
that layers are dynamically unimportant in stellar evolution.

Further evidence of layer instability
is provided by the only numerical experiment on semiconvection
to date, \citet{merry}.  Using the anelastic approximation
for a two dimensional fluid, 
Merryfield studied the onset of overstable oscillations
in a low Prandtl number fluid: the thermal parameters that he used put the 
simulations within the regime of equation (\ref{schwarz_crit}).  
Although motion set in though overstable
oscillations and the compositional gradient 
was seen to mix through the breaking
of internal waves, 
these tended to occur on length scales comparable to the domain, and
layers did not form.
In one of the experiments, an initial condition of two
layers separated by a stable density interface was simulated.  
The interface became 
quickly unstable to internal waves, disrupting the layers and leading 
to well mixed, global thermal convection.  As yet, layers
have not been observed to form, much less be stable, 
in any semiconvective simulation at low Prandtl number.

On the other hand, 
layers have been observed to form in a numerical experiment tailored to
explaining the observations of 
\citet{huplin} and  \citet{fern_89}.  \citet{mole}
considered an incompressible fluid at $Pr = 7$ (water) with vertical salinity
gradient subjected to cooling from above.  Due to Boussinesq symmetry, this is
completely equivalent to heating from below as is done in laboratory
experiments.   The numerical simulation well reproduced the observations
of \citet{fern_89} and confirmed the measurement of a critical Richardson 
number delineating the transition from layer growth by entrainment
to quasi-static convective layers separated by diffusive interfaces.

\subsection{Current Work}

The numerical and theoretical evidence provided thus far verifies the
formation of layers in the high Prandtl number regime appropriate for 
water.  However, the results for low Prandtl number are still inconclusive.
Does a multiple layered structure form in the low Prandtl number regime?
If so, how does the critical Richardson number depend on Prandtl number?

The formation of a second layer above the first requires a
quiescent thermal boundary layer separating the convective 
zone from the static fluid.  In this thermal boundary the slow oscillatory
instability can incubate and grow before convective erosion from below 
becomes significant. 
A smaller Prandtl number means that the viscous boundary layers separating
static from convective zones is smaller than the thermal boundary layer.  
Convective plumes will
tend to decelerate less before reaching the buoyancy interface and be able
to penetrate more energetically.  
Smaller Prandtl number also means smaller
scales for kinetic energy dissipation.  The interfacial waves which are generated
by the convective plumes will tend to break at smaller wavelengths.  The interfacial
waves will then be more effective at mixing the 
composition gradient across the interfaces.  
Both the energetic waves and smaller scale breaking 
tend to suppress the growth
of oscillatory instability in the static fluid.

In the present work I present results of two dimensional numerical simulations
of a fully compressible fluid.  The molecular weight of the medium decreases
upward and the diffusivity of the second species is always small.  
Results for successively decreasing Prandtl number and at different values of
the heat flux are compared.

In section 2 
the equations, choice of parameters and numerical technique are presented.  
The results of the simulations are presented and described in \S 3
and these results are compared to previous mixing length theories
and numerical simulations in \S 4.  In \S 5 
the results are interpreted in the context of stellar models.

\section{Formulation of the problem}
\subsection{Equations and Boundary Conditions}

This work attempts to bridge the divide between terrestrial
double-diffusive convection and stellar semiconvection simulations.
To this end I have, at times, neglected certain processes which
are less relevant to the formation and disruption of layers, though
they are obviously present in realistic stars.  
Radiation pressure tends to decrease the contrast between 
the semiconvective and convective stability thresholds (equations
\ref{ledoux_crit}, \ref{schwarz_crit}) and would correspondingly
mitigate layer formation.  
Although radiation pressure is not negligible in stellar convection, 
for the purposes of this work, it has been neglected.
This serves the added purpose of singling out
differences between astrophysical and terrestrial double diffusion,
since the ideal gas equation of state at constant pressure tends
to the Boussinesq approximation for incompressible fluids.  
Any formation of layers in the present simulations
can be seen as a necessary, but not sufficient condition for
them to form in stars.

The  equations of mass, momentum, energy and
helium concentration conservation which govern
a compressible fluid of hydrogen gas with a 
varying concentration of helium, $Y$, 
are (\citet{spiegel})
\beq
\dt \rho = - \nabla \cdot (\rho {\bf v} )            \label{NS_rho}
\eeq

\beq
\rho \lp \dt {\bf v} + ({\bf v} \cdot {\bf \nabla}) {\bf v} \rp=  
- {\bf \nabla} p +
{\bf g} \rho         +
\nu \rho  \lp\nabla^2 {\bf v} + \frac{1}{3} {\bf \nabla} 
( \nabla \cdot {\bf v} )    \rp
\label{NS_v}
\eeq

\beq
\rho  ~C_v \left ( \dt T + {\bf v} \cdot {\bf \nabla} T \right ) = 
- p ( \nabla \cdot {\bf v} ) 
+  \nabla \cdot (\kappa \nabla T) + 
\nu \rho ( \nabla \cdot ( {\bf v \cdot \nabla v}) - 
\frac{2}{3} (\nabla \cdot {\bf v})^2)
\label{NS_T}
\eeq

\beq
\dt Y  + {\bf v}  \cdot {\bf \nabla} Y  =
\frac{\kappa_{He}}{\rho} \nabla \cdot( {\rho {\bf \nabla} Y})  \label{NS_Y}
\eeq
\beq
p =\frac{ \rho R  T}{\mu} ,
\label{NS_P}
\eeq
where $\nu, \, \kappa, \, \kappa_{He}$ are the constant kinematic viscosity,
thermal diffusivity and helium molecular diffusivity, respectively.  $R$ is the
gas constant, $\mu(Y)$ is the molecular weight for an ionized gas
of hydrogen and helium
\beq
\mu = (2 - \frac{5}{4}Y)^{-1}
\label{eq:mu}
\eeq
and the specific heat at constant volume is
\beq
C_v = \frac{3}{2}\frac{R}{\mu}.
\eeq

Computations are performed on a rectangular box of vertical dimension, $d$
and horizontal dimension $\lambda d$ with
downward directed, constant gravity.  
The initial conditions for the simulations are constant,
negative helium gradient throughout the domain.  In the
bulk of the domain the temperature is chosen such
that the fluid is only marginally unstable to the Schwarzschild
criterion but very stable according to Ledoux.  

An additional heat flux is supplied at the bottom of the domain
and the temperature gradient is continuously set to match it,
exponentially, over a very thin region.  
This has the effect of increasing $\nabla_{ad}$ 
to the point where the fluid is significantly unstable in a thin region 
at the bottom.  This transition region is necessary in order to avoid any numerical
instabilities which arise from impulsive heating.

In all examples, the helium concentration varies from $Y=1$ at the bottom
of the domain linearly to $Y=0$ at the top.  

The side boundaries are periodic while 
the following top and bottom  boundary conditions are used,
\beqan
v_z = \partial_z v_x = 0 & {\rm at} &z = 0, d \\
\partial_z Y = 0 & {\rm at} &z = 0, d \\
\partial_z T = \alpha & {\rm at} &z = d \\
\partial_z T = -(\alpha + F) & {\rm at} &z = 0.
\label{eq:bc}
\eeqan
Physically, the velocity boundary conditions correspond to impenetrable,
stress free boundaries.  The helium boundary conditions are chosen
so that, when convection sets in, there will not be sharp boundary
layers between the convective region and the boundaries.  Such layers
would be computationally expensive to resolve and be irrelevant to the physics.
The initial temperature profile is given by
$T(z,x,t=0) = T_d - \alpha (z-d) +  F z_* e^{-\frac{z}{z_*}} $
where $\alpha$ and $F$ are both positive and $z_* \ll d$.  
$z_*$ is chosen so that the thermal boundary layer at the bottom
is very thin and the temperature at the lower boundary is 
dominated by the $F z_*$ term in the initial condition.
That the heat flux is so low at the top boundary may seem to be cause for concern,
since the driven convection will continue to accumulate energy faster than
it can dissipate it from the top boundary.  In practice, if the temperature
changes significantly at the top boundary, it is due to the fact that a 
convective plume has arrived from the lower boundary.  In such circumstances,
there is no layer formation in the domain (only global thermal convection)
and the simulations are stopped.

\subsection{Numerical Method}

The code is based on one developed by \citet{rubini}
and employs a  mixed spatial scheme:
a psuedo spectral, periodic routine in the horizontal direction and 
$6^{th}$ order compact finite difference in the vertical direction.  
The latter is more appropriate for the boundary conditions under
consideration and the accuracy of the derivatives is very close to 
the spectral accuracy of the horizontal derivatives if double the 
resolution is used.  

The time step for low Mach number compressible simulations 
is limited by the CFL condition for sound waves, 
$\Delta t < \Delta x / c_s$ where $\Delta x$ is the 
mesh spacing and $c_s$ is the sound speed.  
The condition corresponds to the requirement that
sound waves must be tracked across each grid element.
In all cases
considered here, this condition is more restrictive than
the corresponding condition for thermal diffusion, 
$\Delta t < (\Delta x)^2/ \kappa$.  
Since sound waves 
play little role in the convection, it would be 
beneficial to disregard them and their CFL condition as is done in 
the anelastic simulations of \citet{merry}.

The simulations herein are advanced in time by a second order Adams
Bashforth scheme, which is explicit, and were therefore constrained
by the sound CFL condition (see \citet{canuto}).  
The code  employed was
actually developed with the intention of overcoming just this constraint,
having the capability of semi-implicit time-stepping.  
Under appropriate temporal discretization, this allows to jump
the sound CFL condition (see \citet{rubini} for details).  Some
simulations reported herein were reproduced with the semi-implicit
method and the two results were found to agree quite well.  Furthermore, 
after accounting for a factor of 10 speed increase due to the sound
waves and a factor of 2 decrease due to the implicit method, the 
semi-implicit scheme was found to have a net factor of about 5 speed
increase.  Future work will be carried out using this method, but
all of the current results are due to the explicit scheme.  

The results were obtained from the 
SGI Origin 2000 machines,
at Los Alamos and Argonne National Laboratories. Any given 
simulation used from 32 to 96 processors.

Finally note that all of the simulations are fully resolved and do 
not suffer from the need to filter numerical instabilities at
the grid scale.  Though time consuming, this means 
that no numerical diffusion has been introduced thus the
Prandtl and Lewis numbers which are quoted in  table \ref{table:params}
are the actual ones in the simulation.  

\subsection{Physical Parameters}

The dynamical equations are solved by non-dimensionalizing the 
variables in the standard fashion.  All lengths are scaled 
to the height of the domain, $d$, and time to the thermal
diffusion time across the domain $t_{diff} = d^2/\kappa$.  
This naturally leads to the introduction of the Rayleigh
numbers,
\beq
Ra_T \equiv \frac{g d^4 (\nabla_{rad} - \nabla_{ad})}{\kappa \nu H_p}
\label{eq:rat}
\eeq
for heat and
\beq
Ra_{He} \equiv \frac{g d^4 \nabla_{\mu}}{\kappa \nu H_p}
\label{eq:rahe}
\eeq
for helium.  

These parameters arise from the consideration of the linear
instability of a fluid configuration whose initial
state corresponds to a constant molecular weight gradient and 
constant thermal gradient.  Linear and weakly non-linear theory
has been extensively studied with regard to the double diffusive 
problem (see \citet{veronis_65}, \citet{veronis_68}, and the 
thorough review by \citet{turner_85}), but 
the numerical experiments that follow drive the fluid far from
the static equilibrium making both linear and weakly non-linear
discussions  irrelevant. 

For example, even in a thin slot geometry where layering 
is seen to form in laboratory experiments (such as \citet{biello})
and weakly non-linear analysis is greatly simplified, 
\citet{balm_biello}
show that the subcritical nature of the instability prohibits
the solutions from saturating at small amplitude 
and the layering phenomena are not predicted.  Thus far, the 
only theories which successfully  predict layers are the heuristic 
one-dimensional turbulence models of \citet{kerstein} and
the  dynamical mixing-length
models of \citet{balm_etal}. Though the latter
was derived for shear forced layering it may hold insights
for the buoyancy forced layering considered here.
In light of the futility of non-turbulent approaches to the study of layers,
I shall therefore dispense with further discussion of the predictions
of  linear and weakly non-linear theories and concentrate only on the 
highly non-linear problem at hand.

The Prandtl and Lewis numbers, 
already defined above, are the relative measures of viscous
and molecular diffusion to thermal dissipation.
$C_s$ is the non-dimensional
sound speed of the medium and
$H_p = - (\partial \ln p / \partial z)^{-1}$, 
the pressure scale height, has been non-dimensionalized in 
Table \ref{table:params} but not in equations (\ref{eq:rat})-(\ref{eq:rahe}).

The nondimensional parameters are given in Table \ref{table:params},
$\tau = 0.1$ in all simulations and $ Y(t=0) =  1-z/d$  in all except
$VII$ (see below). Astrophysically realistic values have Rayleigh
numbers in the range $10^{12} - 10^{14}$, $\tau \sim 10^{-6}$,
$Pr \sim 10^{-5}$ and $C_s \sim 10^5$.  To perform a completely 
realistic simulation of these parameters is a computationally
impossible task as we would have to resolve the smallest
gradients in the the domain (in this case, due to molecular diffusivity)
and track sound waves across them.  

Keeping in mind that our primary aim is a study of the 
the robustness of layers in the limit of small Prandtl number,
we have chosen fluid parameters in accord with that goal,
limited by computational feasibility and with an ordering relevant to
astrophysics.  Therefore, $ \tau < \sigma << 1 $ preserves the
relative sizes of the gradients of helium, velocity and temperature, 
respectively.  $C_s >> 1$ in all simulations and, correspondingly,
the mach numbers of the flow were also $<1$ (typical maximum 
mach numbers are $0.4$).  

The numerical experiments are organized as follows. 
$I$ and  $II$ 
consider large $(>1)$ Prandtl numbers for varying forcing and domain 
size.  In particular, $I$ is most similar to the Boussinesq regime
and will provide the standard against 
which the possibility of layer formation 
in all of the subsequent experiments, 
which are run at small Prandtl number, will be assessed.  
Simulations $III a,b$ contrast two failed attempts at producing layers
at weak and strong forcing.  
This analysis of this work focuses on the results of  $IV$ and $V$
where a single  layer is formed at low Prandtl
number and migrates upward. In particular, 
the mixing associated with layer growth is studied in detail
in $V$.
In $VI$ an attempt is made to study 
the effect of large aspect ratio on the results of the
previous two simulations.  Finally,
experiment $VII$ is set up as a very stable two layer system with
convection setting in rapidly.  In this setting, one can easily
distinguish the mechanisms of boundary layer erosion, breaking
internal waves and penetrative convection.

\begin{deluxetable}{lccccccc}
\tabletypesize{\scriptsize}
\tablecaption{Numerical Parameters used in the simulations.   \label{table:params}}
\tablewidth{0pt}
\tablehead{
\colhead{Parameter} &
\colhead{$I$} &
\colhead{$II$}&
\colhead{$III \, a,\, b$ }&
\colhead{$IV$}& 
\colhead{$V$}&
\colhead{$VI$}&
\colhead{$VII$\tablenotemark{*} }
}
\startdata
$Pr$  &$3.5$  & $2.0$  &0.25 &0.25  &0.25 &0.25 & 0.25\\

$Ra_{He}\, (\times 10^6)$ \tablenotemark{a}   & $2.5 $ &$0.625 $ 
&$0.625 $ &$2.5 $   &$2.5 $  &$2.5 $  &
$46.0 $\tablenotemark{*}  \\

$Ra_T (\times 10^6)$ \tablenotemark{b}   &$20.0 $  &$4.7 $ 
&$5.0$, $1.0 $&$20.0 $ &$6.0$&$20.0$&$20.0$  \\
$F$   &$10.0$  &$10.0$ 
&$10.0$, $2.0$ &$10.0$&$3.0$  &$10.0$  &$10.0$ \\
$\alpha$   &$0.237$  &$0.325$ 
&$0.175$ &$0.225$ &$0.225$&$0.225$ & $0.225$\\
$H_p$ \tablenotemark{a}   &1.79  & $1.0$
&$4.0$&$2.0$  & $2.0$  & $2.0$  &$2.0$ \\
$C_s \, (\times 10^3)$ \tablenotemark{a}   &$5.0 $ &$1.4 $ 
&$1.0$ &$1.4$  &$1.4 $  &$1.4 $  &$1.4 $   \\

$\lambda$   &$2.0$  &$1.0$ 
& $2.0$&$2.0$&$3.0$ &$4.0$ &$2.0$ \\
nx:nz   &192:320  & 288:576 
&288:576 &288:576 &384:576 &480:480 &288:576  \\

\enddata
\tablenotetext{*}{
This is a two layer simulation with a sharp concentration jump at $z=0.5$.  $Ra_{He}$ is
evaluated at the midplane.}
\tablenotetext{a}{Evaluated at z=d}
\tablenotetext{b}{Evaluated at z=0}

\end{deluxetable}

\section{Results}

The laboratory experiments on salt-water demonstrate that layer
formation is necessarily a transient phenomenon.  The time 
asymptotic state of any layered convective fluid would necessarily
be one of well mixed composition undergoing classical convection.  
For convective layers to be dynamically important to stellar
evolution, not only must they must be demonstrated to form
by some relevant physical mechanism, they must do so 
on time scales short enough for their presence to be felt,
{\em and } they must persist over comparably long time scales.  
In late stages of stellar evolution, such time scales are
of order the evolution time for a star.  Irrespective, such times
are always much longer than the thermal diffusion time, which is
the reference time for all convection calculations.

Due to Rayleigh-Taylor instabilities of the thermal boundary layer at
the bottom of the fluid
the first layer is seen to form rapidly in the laboratory.
The second layer forms from oscillatory
instabilities of the internal boundary layer above the first.  Since the
oscillatory instabilities occur on thermal diffusive time scales, tracing
their complete evolution would be computationally prohibitive 
with the current compressible simulation; to study them over stellar
evolution times would be impossible.  
There is, however, an interesting result from \citet{bruenn}
that double-diffusive instabilities may play a role in the supernova
process, itself,  which could render moot the necessity of considering
large time scales. 
Nonetheless, all of the simulations reported hereafter cease 
at approximately $10 \%$ of a thermal diffusion time; 
too little time to see a second well mixed layer.

To circumvent this problem, 
I shall use phenomena observed during layer formation in
the laboratory as indication that a second layer will form.
Firstly, as a necessary condition, 
the height of the interface separating the convecting fluid
from the nearly stationary flow must be quasi-static,
for experiments (in particular \citet{fern_89}) 
it is seen to grow as $h \sim t^{1/2}$.  
This is tantamount to the restriction that significant 
entrainment across the interface cannot occur, 
or else it would quickly migrate upward.  

Secondly, overstability must been seen to set in above the convecting
flow.  This latter requirement cannot be underestimated, since a 
layer which is quasi-static
over the time scales of the integration hardly can be regarded  as a sufficient
condition for a second layer to form.  In particular, one can envision
a scenario where the lower fluid develops turbulent convection and
the instability in the upper fluid sets in so slowly as to make it
imperceptible over these time scales.  On time scales much longer
than that observed in the laboratory, the convecting region could slowly
entrain fluid from above, avoiding layer formation entirely.

In the descriptions below, 
I shall show when these phenomena are seen or not in the high Prandtl number
simulations and use those results as a guide to study the low Prandtl number
cases.

\subsection{High Prandtl Number}

\subsubsection{Simulation $I$: Layering in a viscous fluid.}

Figure \ref{fig:pr3.5_onset} 
shows the onset of instability at 
early times, for $Pr = 3.5$.  Since the molecular diffusion
is small, the helium fraction, which 
began horizontally constant and decreasing linearly, 
will indicate the convective mixing by the flow.  The color
coding  runs from lightest for $Y=1$ to darkest, $Y=0$, and is chosen
because most of the convective  activity will occur in regions of
high helium concentration; at the bottom.
At $0.715 \% $ of a thermal diffusion time, shown in the figure, the 
Rayleigh-Taylor plumes are clearly evident.  The symmetry
of the flow betrays the symmetry of the initial conditions
which were chosen in order to highlight the relevant mechanisms 
in this illustrative test case.  

\begin{figure}
\epsscale{0.55}
\plotone{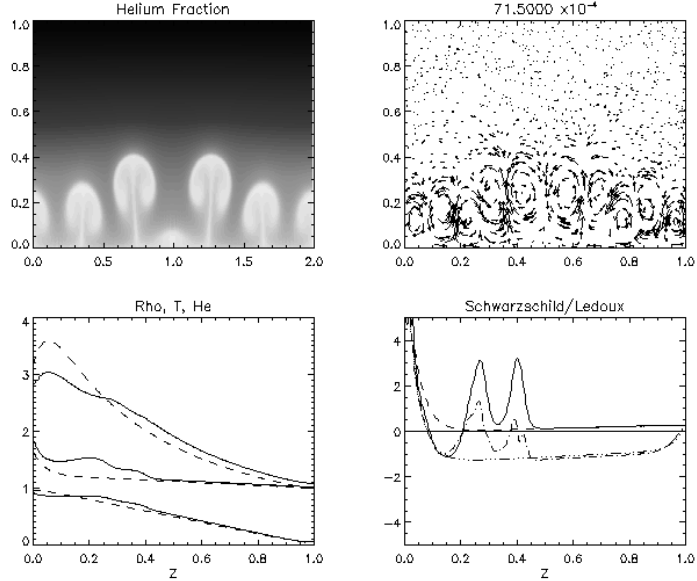}
\caption{Simulation $I$.  
Top left panel: the helium concentration in the domain at $t = .715 \% $
of a thermal diffusion time.  
Top right: the velocity vector field.  The horizontal axis
is actually $x=[0,2]$.
Bottom left: From top to bottom, $\bar{\rho}$, $\bar{T}$, 
$\bar{Y}$.  Solid lines are at $t = .715 \% $, dashed lines are 
initial conditions.
Bottom right: Solid line, $\epsilon_S$.  Dashed line $\epsilon_S(t=0)$.
Dot-dash $\epsilon_L$, three-dot-dash $\epsilon_L(t=0)$.
}
\label{fig:pr3.5_onset}
\end{figure}

In the bottom left panel are plotted the 
horizontal averages of density $(\bar{\rho})$, 
temperature $(\bar{T})$ and helium fraction $(\bar{Y})$ along with
their initial conditions,  respectively, from top to bottom.  
The Rayleigh-Taylor instability is   
an effect of the initial density inversion seen at the bottom of the fluid.
It is also clear that the flow rapidly smooths out that inversion
and layering becomes evident below $z = 0.3$ 
where  $\bar{Y}$ and $\bar{T}$  flatten. 
Moreover, owing to the relative diffusivities, the
transition from mixed fluid (little or no gradient) at the bottom to constant 
gradient at the top is much sharper for  $\bar{Y}$ than for $\bar{T}$.

The final panel shows the local value of $\epsilon_S =
\nabla - \nabla_{ad}$ (Schwarzschild
criterion) and $\epsilon_L =
\nabla - \nabla_{Led}$ (Ledoux criterion) 
determined using the horizontally averaged thermodynamic quantities and
compared to their respective initial values.   Initially $\epsilon_L$ is
everywhere negative except for the temperature transition region
at the bottom where it rises to coincide with the maximum value
of $\epsilon_S$ (since $\nabla_{\mu}=0$ near the bottom boundary).  
Conversely, $\epsilon_S$ is only slightly supercritical in the bulk
of the domain and also rises at the lower boundary.  
Though the region above the transition layer is stable, the momentum
of the  Rayleigh-Taylor plumes allow them to penetrate
deeply and consequently homogenize the temperature and helium.

Figure \ref{fig:pr3.5_oscil} 
shows a well developed, almost fully
mixed lower layer at $t=2.74 \times 10^{-2}$.  The
figure does not reval the steady internal wave of 
wavelength equal to the domain width and height, $h$, 
$ \sim 0.25$ which is also present.  
In the horizontal averages there is a flat distribution of helium
and almost flat distribution of temperature below $z=0.4$.
The internal helium boundary layer, extending $z=[0.45,0.5]$, is of 
thickness $\delta_{He} \sim 0.05$ whereas the internal temperature
boundary layer, $z=[0.4,0.55]$, is of thickness $\delta_{T} \sim 0.15$.
This is consistent with the predictions of \citet{spruit} that
the relative thickness of diffusive internal layers of temperature
and helium in layered convection
should be of order 
$d_{He} / d_{T} \sim \tau^{1/2} (\sim 1/3$ for all of the results herein).  
Such estimates are straightforward to derive by arguing 
that the convection homogenizes the interior and a convective
turnover time sets the scale over which which material (or heat)
can diffuse out of a fluid blob at an internal layer
(or boundary layer, see \citet{shraiman}).  

A sign of efficient convection is that the fluid be almost adiabatic, and
this is evidenced by the last panel of figure \ref{fig:pr3.5_oscil} where
$\epsilon_T \sim \epsilon_S \sim 0$
over $z=[0.15,0.4]$.
Furthermore, the 
internal thermal transition layer and the entire region
above it is unstable to oscillations yet stable to overturning convection.
In fact, a weak two cell oscillatory flow 
is seen in the velocity vector plot and
will be elaborated upon below.

\begin{figure}
\epsscale{0.55}
\plotone{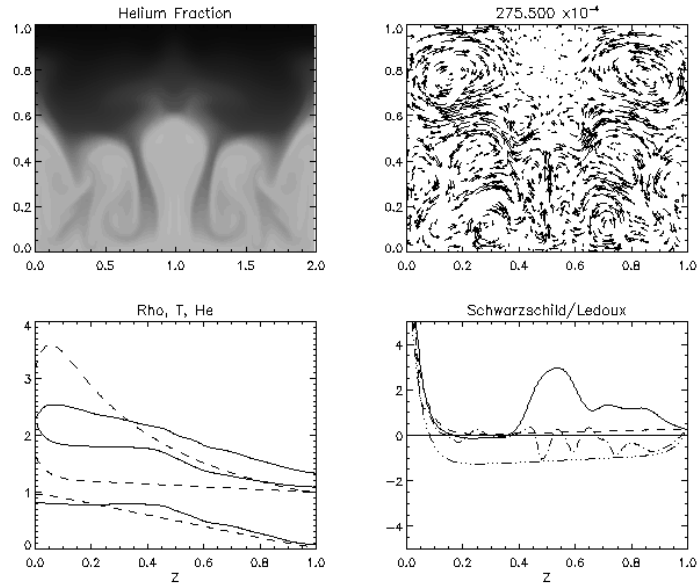}
\caption{Simulation $I$.
Same as figure \ref{fig:pr3.5_onset}
except at $t=2.755 \%$ of
a thermal diffusion time.  The lower half of the domain is
a well homogenized layer, as is seen in the helium
concentration and mean field plots.  The 
internal temperature transition layer is much thicker than the helium
interfacial layer.  The Schwarzschild/Ledoux criterion
predict that the thermal boundary layer is 
unstable to oscillations but stable to convection.
Weak oscillatory rolls are seen in the 
upper half domain in the velocity vector plot.
}
\label{fig:pr3.5_oscil}
\end{figure}

Three mechanisms contribute to mixing helium  across the interface, 
the first being molecular diffusion enhanced by the sharp gradients there.
From figure \ref{fig:pr3.5_oscil}
it is also clear that
there are significant regions where low helium concentration fluid 
is down-welled into the convection zone.  Such regions are a remnant
of both the original Rayleigh-Taylor plumes and the large amplitude
interfacial waves which they excite.  As time proceeds the 
lower layer becomes more well mixed, the interface sharpens and
mixing by down-welling becomes less relevant.
Finally, the scouring of the helium internal layer is a 
major source of vertical helium transport at all times.

The layer height is determined from $\bar{Y}$ by 
finding the height where $|\partial \ln \mu(\bar{Y})/\partial z|$ is maximum.  
Plotted in the first panel of 
figure \ref{fig:pr3.5_layer_ri} is the interface height as a function
of time.  After an initial growth phase, and except for internal
wave oscillations, the layer height is seen to remain relatively constant
with time.  This implies that the sum of the interfacial mixing 
mechanisms do not contribute enough  to destabilize the interface.
Therefore the necessary condition for the formation of 
a second layer, a quasi-static interface, is satisfied.

\begin{figure}
\epsscale{0.55}
\plotone{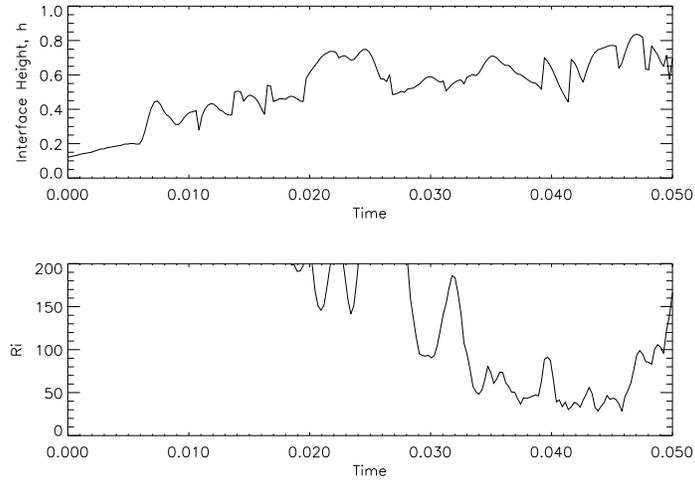}
\caption{Top: interface height as a function of time for simulation $I$.
Bottom: Richardson number versus time.  Until a layer is well
developed, the Richardson number is not relevant, so we can
disregard the high values of $Ri$ at early times.
}
\label{fig:pr3.5_layer_ri}
\end{figure}

The third mixing mechanism, internal layer entrainment,
has the added effect of driving oscillations
in the upper fluid in the following manner.
As the interfacial wave rises, it drives an upward flow
directly above the wave crest.  
Since the flow above the interface rises on hot plumes from below, this
behavior is often referred to as the buoyancy coupling of two layers. 
When the wave turns downward, the flow in 
the upper domain continues upward and, through shear drag,
is able to strip some of the helium 
rich fluid from the interface.  Locally, fluid that has been stripped 
has greater helium concentration than its environment, 
yet due to its momentum it continues upward.  
Before being able to overturn completely, 
the density excess decelerates the flow and the roll turns around:
this is the oscillatory instability of the  upper fluid.

The helium rich, upward flowing fluid at $(x,z) = (1.0,0.75)$
in figure 
\ref{fig:pr3.5_oscil} is a product of this internal layer stripping.  
Plotted in figure \ref{fig:pr3.5_vel} 
is the time trace of
the vertical velocity field
and $500$ times the helium deviation from its initial value at this point.
The solid line clearly shows growing oscillations with period $\approx 1\%$ of
a diffusion time and a longer growth time.  The helium deviation 
is slightly out of phase with the velocity, which drives the oscillations.
Parenthetically note that the ripples on the velocity trace
clear at $t=2\%$ and elsewhere are due to sound waves.

\begin{figure}
\epsscale{0.55}
\plotone{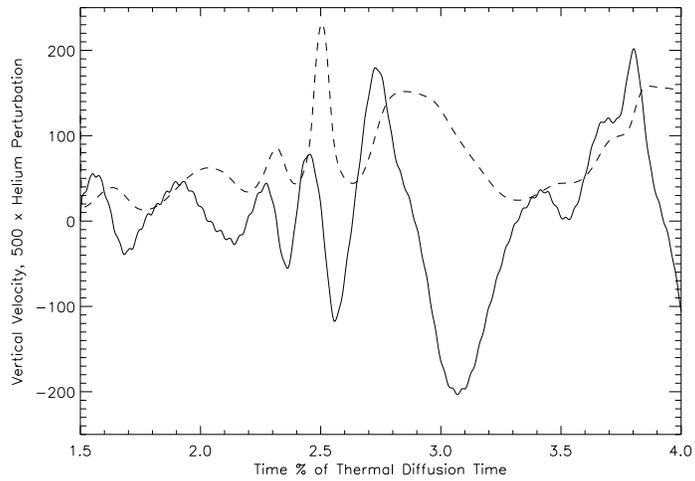}
\caption{
Solid line: Vertical velocity field versus time 
in simulation $I$ at $(x,z) = (1.0,0.75)$.
Dashed line: $500 \times $ helium deviation at that point.
}
\label{fig:pr3.5_vel}
\end{figure}

Though a layer has not yet fully developed by the end of the simulation $I$,
one is obviously in the process of doing so.  
Since it contains both phenomena which are observed in laboratory
layer formation, I conclude that it would likely form one if allowed to continue.

\subsection{Simulation $II$: A vortex/shear layer.}

\begin{figure}
\epsscale{0.55}
\plotone{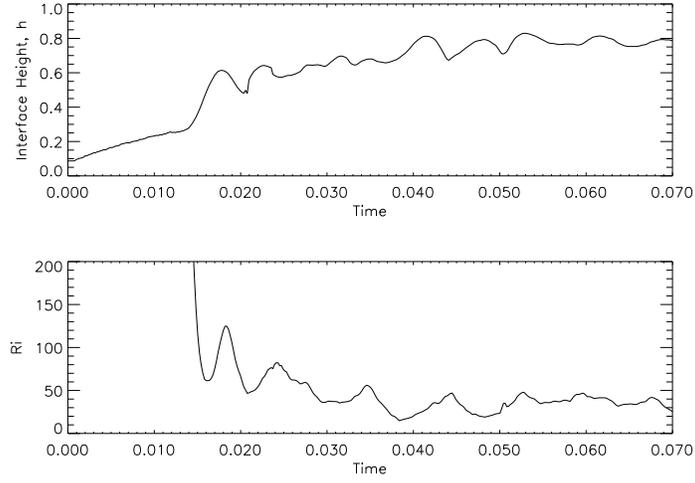}
\caption{
Top: interface height as a function of time for simulation $II$.
Bottom: Richardson number versus time.  
}
\label{fig:pr2.0_ri}
\end{figure}

The important differences of the second simulation are the lower
aspect ratio, weaker forcing and  initial conditions.
For this run the initial velocity is set to a 
random spectrum of very low amplitude 
vortical waves with a  wavenumber cut off at an intermediate length scale
$(k_{max} \sim 20)$.

Again, the instability sets in through  isolated, randomly
spaced Rayleigh-Taylor  plumes rising from the lower boundary.
However, soon the flow becomes coherent and 
mixes the helium well in the lower half of the domain.
At $t=0.07$  figure \ref{fig:pr2.0_ri} 
shows that 
the interfacial layer height has saturated at $z=0.8$.

The flow in figure \ref{fig:pr2.0_soliton} 
is exceedingly elegant in
its simplicity and merits some elaboration.  The stable
concentration jump is clear at $z=0.75$ in that $\epsilon_L < 0$, while
$\epsilon_S >0 $ indicates that overstable oscillations should
eventually set in.  
Moreover, an interfacial wave of the size of the domain 
propagates to the right.
In the bulk of the well mixed  region a strong
shear layer has spontaneously arisen, reminiscent
of the shear flow seen at high Rayleigh number
Rayleigh-B\'{e}nard convection (though much less turbulent).

The shear at the bottom boundary aligns the
flow so that heat fluxes into the 
fluid through one coherent thermal,
located at $x \sim 0.8$ in the figure. 
Being carried by the shear, this thermal propagates leftward.

The entire shear layer is not stationary, rather it
oscillates between closed flow lines and
a single closed vortex, the eye of which is reforming at
$(x,z) = (0.15,0.25)$.  
\citet{merry} produced a coherent structure
in a weakly forced simulation, though in that
case the vortex was a double gyre.
In dynamical systems literature, 
an oscillating jet/vortex flow is referred to as separatrix reconnection, 
and takes place when a propagating wave is modulated
at another frequency.  In this case, the second period
is provided by the leftward propagating thermal at the base of
the flow which interacts with the rightward propagating interfacial wave above.

\begin{figure}
\epsscale{0.55}
\plotone{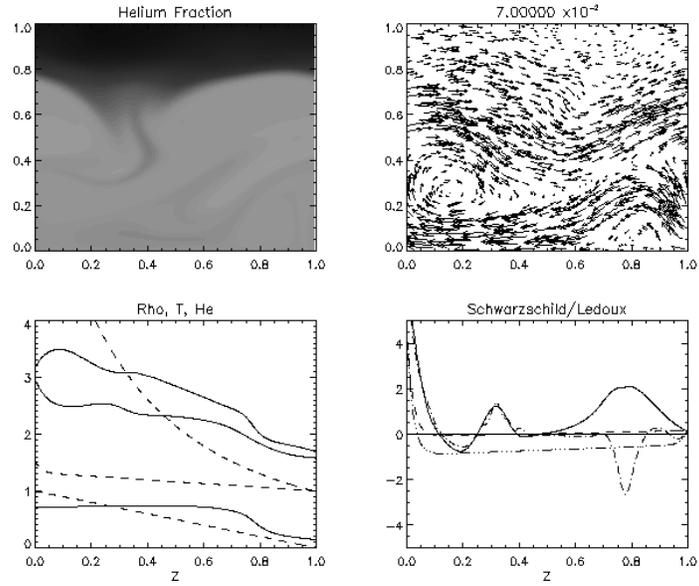}
\caption{A thick layer with sharp density jump
is clear seen in both the $Y$ concentration and Ledoux Criterion 
in simulation $II$.  
The strong strong shear layer and traveling coherent vortex form spontaneously
and are long lived.  Helium transport is enhanced by
separatrix reconnection as is in the low helium at the trough of
the interfacial wave in the upper left panel.
}
\label{fig:pr2.0_soliton}
\end{figure}

Jets of collimated flow, like the
one below the interface in figure $3.6$
are notoriously stable to advective transport across them,
and one would expect  molecular diffusivity to be
the primary mode of helium mixing across the interface.
However, separatrix reconnection provides a mixing
channel so that, even in this simple flow, it is able to
entrain helium poor fluid through the trough of the interfacial
wave (particularly evident in the plot of helium fraction).

Finally, take note of the small knee in the temperature 
average and the corresponding peak in $\epsilon_S, \epsilon_L$.
This is the imprint of the coherent thermal which, owing to 
the vortex and shear, mixes with the rest of the flow through
discreet release events.

Though performed at high Prandtl number, this simulation holds an 
important lesson for astrophysical layering.  Even in a
weakly forced scenario where the interfacial waves are
very low amplitude, mixing by separatrix reconnection will
provide much greater cross interface helium transport 
than molecular diffusion  enhanced by the sharp gradient there.

\clearpage
\subsection{Low Prandtl Number}

\subsubsection{Simulations $III$: No layering}

Simulations $III a,b$ are presented as two cases in which, 
according to the analytic theory described in 
\S \ref{sec:compare}, interfaces should form.
In both cases layered convection does not occur, but for different reasons.

In $III a$ the stabilizing compositional gradient is not sufficient to 
impede the rising Rayleigh-Taylor plumes. 
Consequently, they hit the top boundary before turning back and eventually
coalescing.  By $t=0.061$ in figure \ref{fig:pr.25_nolayer}
 the helium is already well mixed and the convection is global.

\begin{figure}
\epsscale{0.55}
\plotone{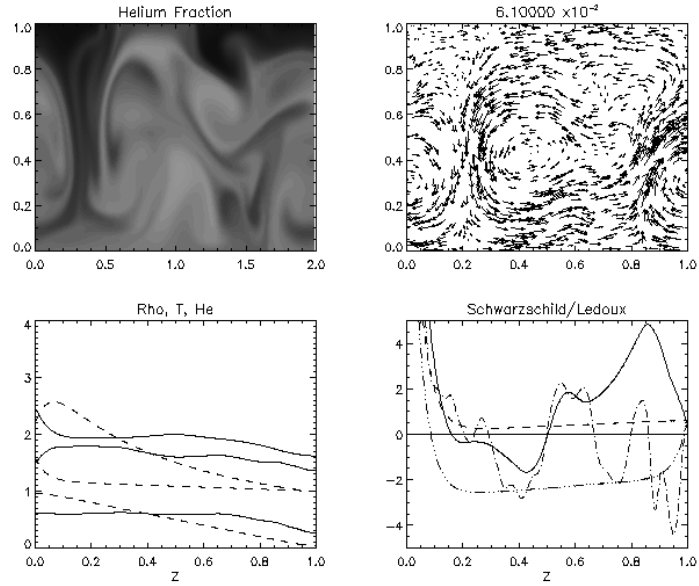}
\caption{
At $t = 0.061 $ the convection in $III a$ 
has effectively mixed the whole fluid.
}
\label{fig:pr.25_nolayer}
\end{figure}

A relation for the quasi-static interfacial layer height in terms
of the heat flux,  the initial buoyancy frequency of the
domain and empirically determined factors
was given by \citet{fern_89} and will be discussed at length below.
Suffice for the moment to say that the critical height varies as 
the square root of the heat flux at the lower boundary 
and is thus proportional to $\sqrt{Ra_T}$ there.
Guided by this and the previous failure,  a heat flux one fifth that of
$III a$ is chosen, with the expectation that a layer should saturate 
within  the height of the domain.
However, by $t= 0.07$ in $III b$ the lower thermal boundary layer has diffused
to $z=0.5$ though convection has yet to set in.
Correspondingly, the whole region $z<0.5$ has become Ledoux unstable.
It is reasonable to conclude that
the growth rate of the plumes is too low
to  observe convection before the Ledoux unstable region 
encompasses the entire computational domain.  Global convection would thus
ensue.

This is an interesting result, but
probably less relevant for astrophysics on two counts.  First, astrophysically
high Rayleigh numbers yield correspondingly large growth rates.  Second,
the simulation was initialized with a small velocity field; 
more realistic in nature would be 
a spectrum of finite amplitude gravity waves, allowing to 
avoid the, somewhat long, linear growth phase.

\subsubsection{Simulations 
$IV$, $V$, $VI$: Layer formation and interface migration}

\begin{figure}
\epsscale{0.55}
\plotone{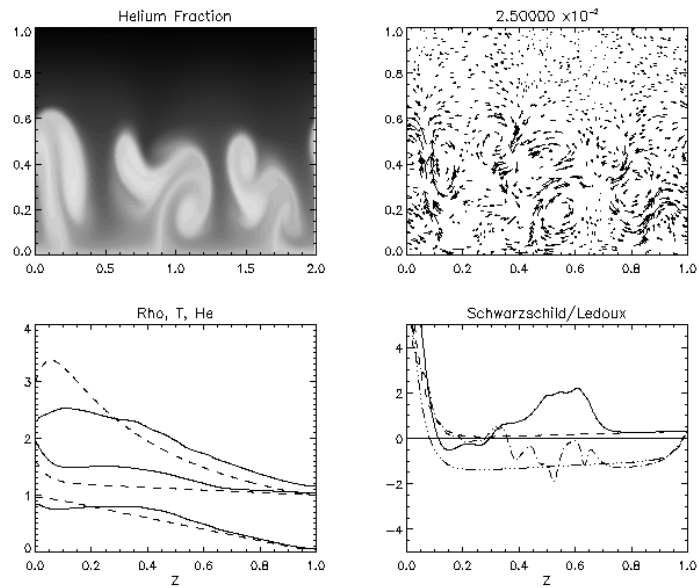}
\caption{At $t=0.025$ in simulation $IV$ the plumes have
mostly stopped at the midplane through a combination
of buoyancy contrast and enhanced drag.  
Parts of the plumes which remain buoyant continue upward.
}
\label{fig:pr.25_fl10_1}
\end{figure}

Stars have  both small Prandtl number and exceedingly
large Rayleigh numbers and, to the limit of computational feasibility,
these conditions are reproduced in 
simulations $IV, \,\,V$ and $VI$.

\begin{figure}
\epsscale{0.5}
\plotone{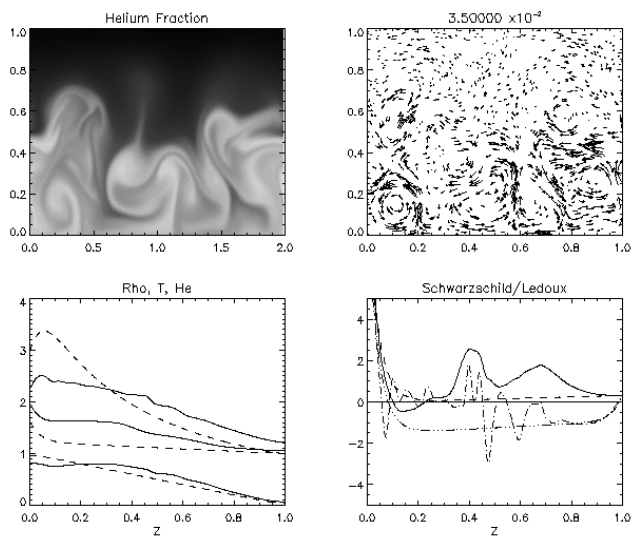}
\caption{At $t=0.035$ in $IV$ the plumes have overshot their equilibrium
height and fall.  A lower layer, clearly evident in the horizontal
mean fields, begins to form.  The thin helium rich filament
above the mixing region has been stripped from a plume which 
overshot. 
}
\label{fig:pr.25_fl10_2}
\end{figure}

\begin{figure}
\epsscale{0.5}
\plotone{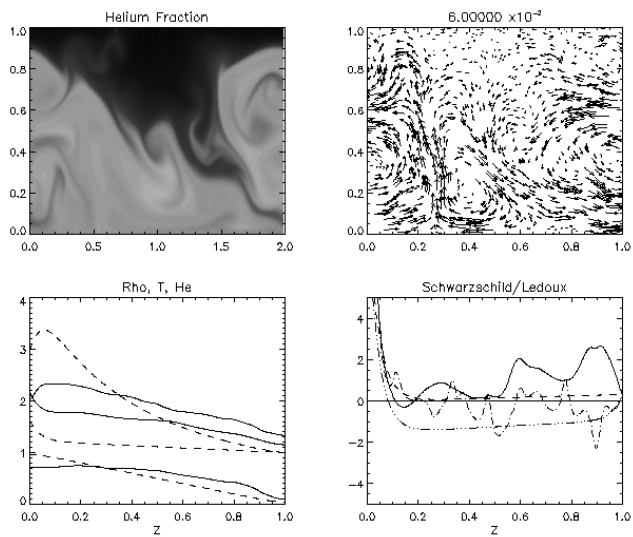}
\caption{Though the flow is highly disordered in $IV$, a large internal
wave dominates the mixing through down-welling.  The mean fields
and $\epsilon$ give no hint of the highly distorted interface.
}
\label{fig:pr.25_fl10_3}
\end{figure}

\begin{figure}
\epsscale{0.55}
\plotone{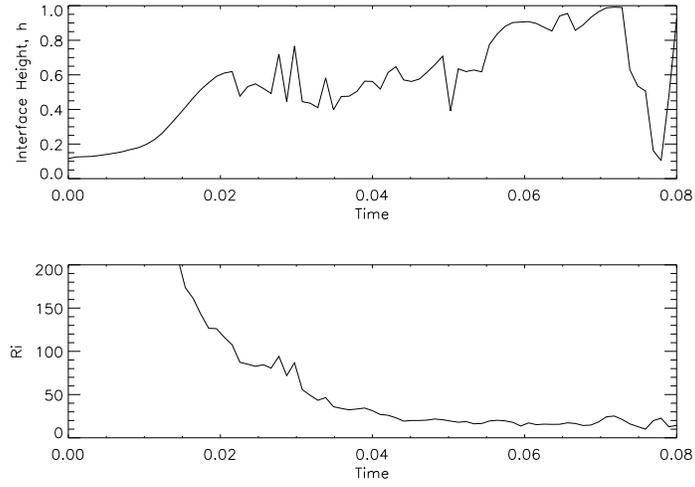}
\caption{Interface height and Richardson number for $IV$.
}
\label{fig:pr.25_fl10_layer}
\end{figure}

Figures \ref{fig:pr.25_fl10_1} - \ref{fig:pr.25_fl10_3}
are the $t=0.025,$ $0.035$ and $0.06$ snapshots of simulation $IV$ in
which a small initial velocity spectrum is used to seed the
Rayleigh-Taylor instability.
By $t=0.025$ the plumes have reached $z=0.5$ and slowed significantly.  
Notice, though, that smaller plumes with a large helium contrast
are still able to rise from the interface of the main plumes;
this  plume splitting is seen at the center of the $Y$ 
concentration plot.
Though mostly having stopped near the midplane the
plumes are still slightly buoyant, therefore the
primary mechanism slowing them must be their drag.  

Even though viscosity is weak, it is well known
that drag can be significantly 
enhanced by the adiabatic expansion  of rising plumes,
the entrainment of fluid into the plumes and
the breakup of the plume tip. 
Since it is clear that
parts of the plumes remain buoyant after breakup and continue
to penetrate up to  $z=0.7$
then, necessarily, buoyancy and thermal 
diffusion out of the plumes play
a  smaller role in their deceleration up to the midplane.

In figure \ref{fig:pr.25_fl10_2}
the plumes have turned around and begin to resemble
convecting cells mixing the bottom half of the domain.
However, during the deep penetration, 
there is significant mixing of 
helium rich fluid  up to $z=0.8$, the shadow of which are
the faint plumes seen in the upper half plane.
Like the scoured interfacial layer in simulation $I$, 
these faint plumes have a $Y$ greater than their surroundings
and can be expected to descend.  Also like their counterparts,
these plumes are in a region which is apparently
oscillatorily unstable.  
However, a clear interface 
has yet to form above the convecting region
and no oscillatory instability is apparent for $z>0.5$
at $t=0.035$.

From interface formation to $t=0.06$ in figure \ref{fig:pr.25_fl10_3}
the flow becomes turbulent, but with a persistent large scale component.
In turn, these few large vortices and plumes drive an interfacial
wave of wavelength approximately equal to the domain length. 
At this point $\epsilon_L$, $\epsilon_S$ are irrelevant measures
of convective stability as the interface is clearly not horizontal.
Unlike the high Prandtl number case, this interfacial wave is of very 
large amplitude and all of the mixing mechanisms thus far discussed
take place across it.
Needless to say, molecular diffusion enhanced by the interfacial gradient
plays a relatively small role in the total transport.

Most obvious from figure  \ref{fig:pr.25_fl10_3}
is the strong down-well obliquely crossing the domain and upon which
a helium rich internal wave is about to break.  This is a
more dramatic version of the separatrix reconnection demonstrated in
$II$.  Traces of previous mixing events are seen as helium poor filaments
in the otherwise mixed region.  
Less obviously, 
the upper left of the thick helium boundary layer 
is experiencing scouring by a flow along the interface
(around $(x,z) =(0.5,0.9)$).
Moreover, a down-flow near the mouth of the down-well (at the
center of the domain) is clearly shearing two helium lamina, though
where they will mix is not clear.  Irrespective, 
this flow is violently mixing the domain and
by $t=0.07$ the interfacial wave collides with the upper 
boundary so that no second layer can be formed.

Interface height is plotted against time in figure 
\ref{fig:pr.25_fl10_layer}.  Except for the last fraction of a diffusion
time  ($\sim 10^{-3}$),  the interface
compares well with that of figure \ref{fig:pr3.5_layer_ri}.  
The spikes betray the large amplitude internal waves on the interface and the
collapse of the interface near the end reflects the overturning convection
which sets in at that time.

\begin{figure}
\epsscale{0.85}
\plotone{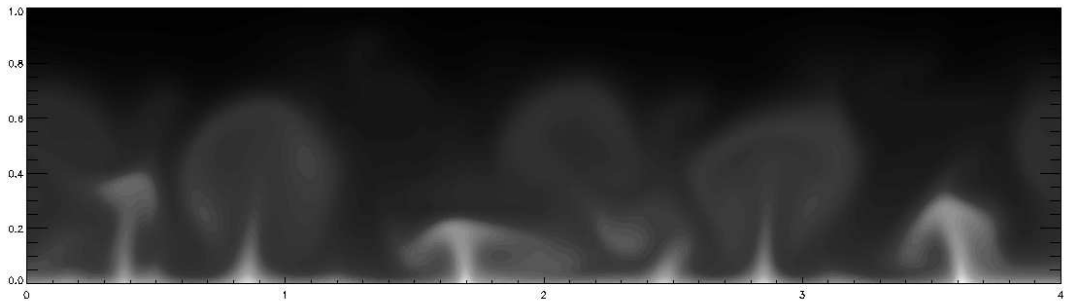}
\caption{Temperature plot of $VI$ at the $t = 0.03$ showing
isolated thermals emanating from  the bottom boundary.  In the helium field
(not shown) the plumes have begun to form an interface at this time.
}
\label{fig:wide}
\end{figure}

Each of the dynamic mixing mechanisms have at their source the
fact that the amplitude of the interfacial wave is very large, comparable
to its wavelength.  This, in turn, may be a consequence of  both
the fact that the flow under consideration is 2-dimensional and 
that the aspect ratio of the domain is small,  $\lambda=2$.  
In  3-d simulations this inverse cascade would not occur 
while, since rolls will not likely have wide aspect ratio themselves,
it would be irrelevant for large aspect ratio 2-d boxes.

In simulation $VI$ an attempt is made to study the effect of
large aspect ratio on, in particular, the amplitude of the interfacial
wave.  Unfortunately, the simulation lost resolution and had to 
be ended before an interface was well developed.  
Nonetheless it
is evident from the temperature field 
shown in figure \ref{fig:wide} that rising thermals are 
well spaced along the bottom boundary
and there is some hope that, should
many weak thermals persist, the flow would not produce 
such large scale structures
particularly, the catastrophic internal wave.

\begin{figure}
\epsscale{0.5}
\plotone{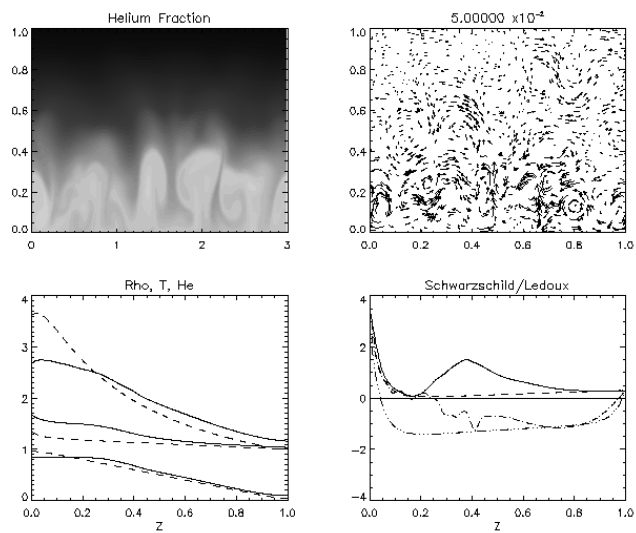}
\caption{After $t=0.05$ 
in simulation $V$ a sharp layer has formed  below $z=0.35$,
where the convection is less vigorous than in previous simulations.
Nontheless, the convection generates narrow structures which
account for most of the mixing with the static fluid above.
}
\label{fig:V_1}
\end{figure}

The most promising case for layer formation
is provided by simulation $V$.  Akin to $IV$, 
it  has lower heat flux at the bottom boundary and
larger aspect ration ($\lambda = 3$), the objective being
that the layer height should saturate at smaller $z$ and
that the convection should not be dominated by large 
structures.  
Figure \ref{fig:V_1} shows that this is indeed the case since
by $t=0.05$ the interface has reached a quasistationary height of
$z=0.35$.  Notice that the time is much later than in 
previous simulations since the flow is much weaker.  
The thermodynamic  criteria are favorable to onset of
oscillations just above the interface and already
at this early time such oscillations are seen in the
fluid above. 
Despite weak convection in the lower layer,
the interface in figure \ref{fig:V_1} is strikingly irregular.  
Though the interfacial waves remain small, mixing is 
again dominated by large structures being engulfed
between the plumes  below the interface.  

By $t=0.096$ the interface has continued to migrate upward,
and the density jump is  below $z=0.6$ in figure 
\ref{fig:V_2}.  Interfacial scouring due to oscillatory
convection in the upper fluid creates filaments of helium
rich fluid above the interface and mixing below remains
dominated by large blobs of fluid between the plumes.  

\begin{figure}
\epsscale{0.5}
\plotone{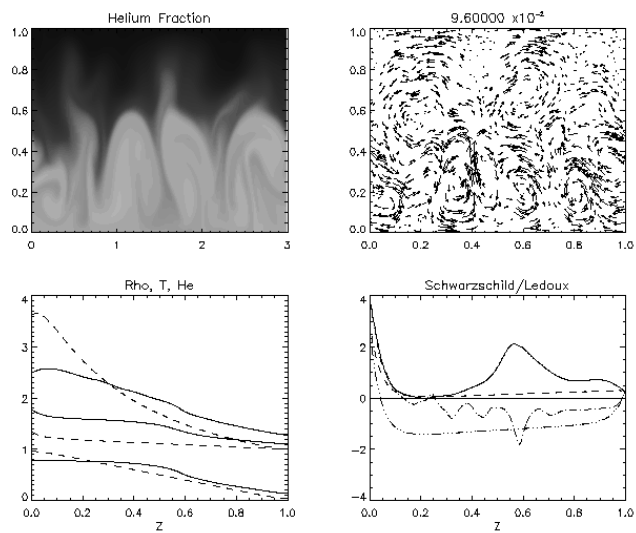}
\caption{At $t=0.096$ the layer has migrated to $z \sim 0.5$
and interfacial waves are quite large.  Entrainment at wave
crests continues to be the main source of mixing.  By now, 
weak oscillatory convection is taking place in the upper half of the fluid
and causing helium to be scoured upward from the layer interface.
}
\label{fig:V_2}
\end{figure}

Large plumes have penetrated to $z=0.8$ by $t=0.136$ (figure \ref{fig:V_3})
effectively destroying the oscillatory flow which had been growing
there.  Two plumes dominate the flow at this time and are clearly
able to entrain large, helium poor fluid elements downward into the convection
as is seen at $(0.6,0.3)$ in the figure.  Interfacial splashing, evident
in a thin filament above this blob, overtakes scouring as the principal
mechanism for helium transport above the  interface.  

Though the vestiges of a layer are evident in the mean helium field,  
the interfacial waves have grown large enough to nearly reach
the top boundary, suggesting that a second layer will
not form before the interface grows to encompass the whole fluid.
Such waves, though less marked, are clear in figure \ref{fig:ri_h_V}.
Nonetheless, this is the only case discussed thus far where the theoretical
prediction of \citet{fern_89},
$h \sim \sqrt{t}$, provides an excellent fit.
The curve $h= 1.8 \sqrt{t}$ agrees well with the data after $t=0.01$
and despite the growing interfacial waves at late times.  Furthermore
this fit yields an effective diffusion coefficient across the interface of
$\tau_{eff} = 1.62$: it will be shown subsequently that this greatly
exceeds the mixing provided in existing models.

\begin{figure}
\epsscale{0.5}
\plotone{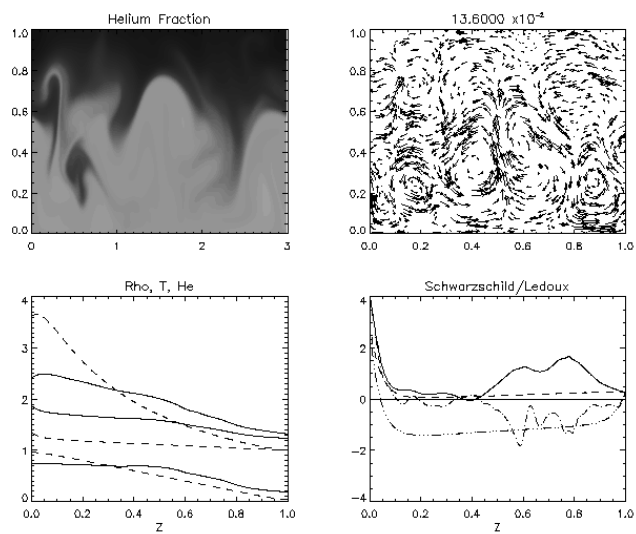}
\caption{At the end of $V$, at $t=0.0136$
a large blob of helium poor fluid is being entrained into the lower layer.
Though the thermodynamic criteria remain favorable for a second layer
to form and there persists a 
helium gradient  in the top $z=0.4$ of the domain, it seems unlikely
that oscillatory convection will be able to overcome the entrainment 
from the interface.
}
\label{fig:V_3}
\end{figure}

Even in this most weakly forced example where the aspect ratio
of the convecting fluid remains greater than $6$ for most of the
simulation, there persists an inverse cascade to large scale structures.
Figure \ref{fig:plume} shows a space time plot of the
temperature perturbation on the bottom boundary.  
Hot spots, or thermals, along
this boundary are seen as bright filaments in this plot; from here plumes 
rise and convection is driven.  It is clear that the main plume dynamic
is one of coalescence for, even though approximately $10$ thermals
arise at the onset of convection, only two large thermals 
are apparent by the end.  Furthermore, weak plumes are continually
formed and incorporated into the stronger ones since the convection
tends to direct flow toward strong plumes on the boundary. 
Two plumes implies four large scale rolls and since at
$t=0.136$, $h \sim 0.7$ the aspect ratio of each roll is $\sim 1.1$,
less than, but not atypical of large structure in regular
Rayleigh-Benard convection.  
One must therefore conclude that the aspect ratio of layered
convection will remain $\approx 1$ irrespective of the layer height.

\begin{figure}
\epsscale{0.55}
\plotone{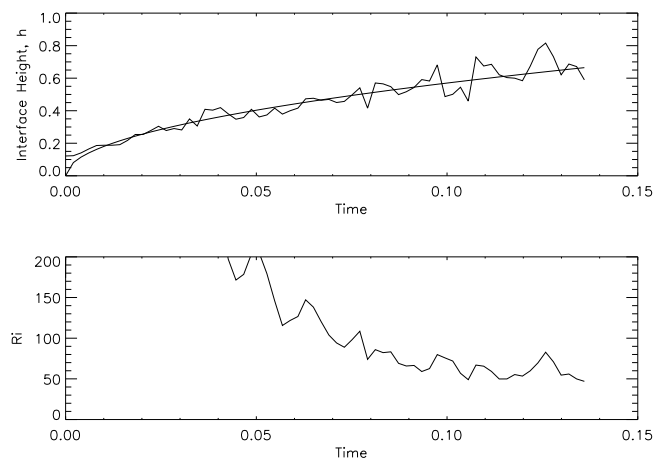}
\caption{Layer height and Richardson number versus time for
$V$.  A fit of $h = 1.8 \sqrt{t} \,\,$ is included and yields
a effective diffusion coefficient of $\approx 1.62$ (see
figure \ref{fig:diff_coeff}).
}
\label{fig:ri_h_V}
\end{figure}

\begin{figure}
\epsscale{0.5}
\plotone{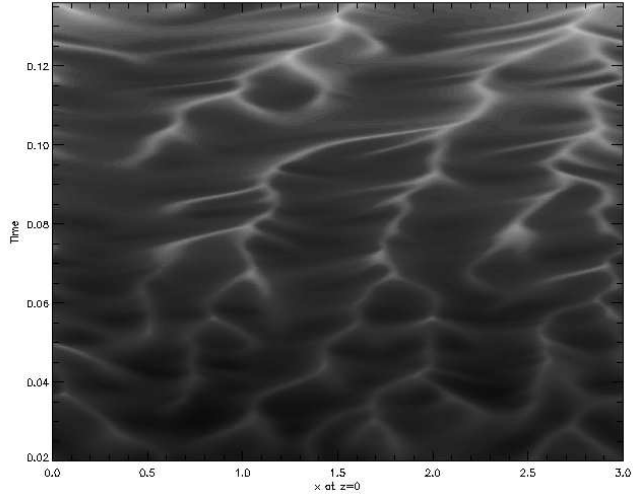}
\caption{Space-time plot of temperature perturbation
evaluated along the bottom boundary in simulation $V$.  There are clearly
several hotspots at early times which eventually coalesce and
become plumes.  By the end (top of the figure) only two, very
strong plumes remain.
}
\label{fig:plume}
\end{figure}

Therefore, as to the the question of whether a second layer forms,
the results for low Prandtl number are inconclusive
but not promising.
Since the large amplitude internal wave seen in $IV$ 
and $V$  drive instabilities
along the interface and are a source for much cross interface mixing
it remains possible that any layer can continually entrain fluid
from above before oscillatory convection becomes well developed.
Whether this is
simply due to the strong forcing or to the relatively small 
aspect ratio under consideration remains to be clarified.
If due to the latter, then
it can be argued that in a stellar context where aspect ratios
are large, such a large interfacial wave will not exist.

If, in fact, the fluid is forced too strongly in $IV$ and $V$
then there still remain several possibilities.  
It is clear that the convection 
continues to inverse cascade to larger scales at each stage, 
so that plumes within a layer are always spaced at 
a distance about twice the layer height.
The plumes generate the interfacial waves reaching amplitudes of
significant fractions of the layer thickness (particularly
clear in figure \ref{fig:V_3})
which may persist, constantly entraining
fluid from above in broad down-flows.  
In such a circumstance an oscillatory instability would be unable
to set in above the interface and eventually a state of global, classical
convection would exist.

On the other hand, the interface may continue to 
migrate upward, gradually stabilizing 
until the plumes flatten along the buoyancy jump at the interface
and  are no longer able to excite large waves there.  This 
mechanism  occurs in the laboratory and is the motivation for
introducing the Richardson number (equation \ref{eq:richardson}).
A low value of $Ri$ implies that the interface should migrate upward
and \citet{fern_89} measured a critical value for this behavior
in the laboratory.  
The bottom panels in figures \ref{fig:pr3.5_layer_ri}, $3.5$,
\ref{fig:pr.25_fl10_layer} and \ref{fig:ri_h_V}
trace the interfacial Richardson number with time for simulations
$I$, $II$, $IV$ and $V$, respectively.  
That $Ri$ is smallest in case $IV$ is consistent with the observed
layer migration; in fact,
in case $IV$ it is just within the range of the critical number
measured in the laboratory 
experiments of  \citet{fern_89} where $Ri_{crit} \sim 2 - 30$.  

Throughout simulation $V$, however,  the Richardson number is greater than 
$\sim 50$, well above the critical value measured by \citet{fern_89}.
However, the interface continually migrates upward and interfacial
waves increase with time, as opposed to decreasing.  
Compare this with simulation $I$ (figure \ref{fig:pr3.5_layer_ri})
where $Ri < 50 $ for a short
period of time yet  the interface does not migrate appreciably.
Such observations can be most readily explained if the interfacial
Richardson number at late times is below the critical value and thus
provide strong evidence that the critical interfacial
Richardson number should increase with decreasing  Prandtl number.

\section{Low Prandtl Number Convection Impinging on a Stable
Compositional Interface.}

In order to isolate the mixing experienced at a stable interface
from the question of the formation of that interface, 
simulation $VII$ is initialized with $Y=0$ in the top
half of the domain, $Y=1$ in the bottom and a thin
($\sim .05$) transition region between.  

By $t=0.014$, (figure \ref{fig:2layer_1}) the convection has set
in throughout the lower layer.  
Though the density jump still stabilizes the 
interface, waves are apparent.  
Again, mixing takes place 
at the troughs of these waves and in the helium plot there are 
two significant helium rich structures penetrating into the upper
layer.  Some interfacial scouring has occured, but is
apparently much weaker than the wave breaking.
Although
$\epsilon_S >0$ just above the interface,
there is yet no motion in the upper layer beyond
its interaction with the interfacial wave.

\begin{figure}
\epsscale{0.55}
\plotone{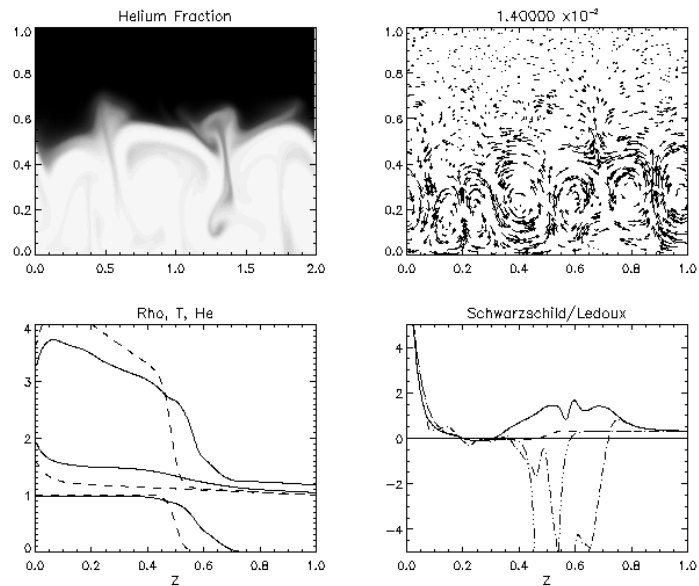}
\caption{Convection has set in throughout the lower
layer by $t=0.014$ in simulation $VII$.  Interfacial
waves are still small, but on breaking they mix
significant amounts of helium into the upper layer.
Scouring is apparent just above the interface at $x=1.8$.
}
\label{fig:2layer_1}
\end{figure}

By $t=.027$
the interfacial wave breaking is very significant and
a large mixing plume is seen in figure \ref{fig:2layer_2}.
The concentration jump has barely been eroded and the
interfacial wave is still of moderate height ($\sim 0.07$)
yet its self interaction is able to generate large structures, 
much like the interaction of two solitary waves on 
the surface of water.  

In the upper half of the domain, 
not only is flow visible around the helium plume, but also ahead of
it, resembling the upstream field of a weak vortex
(notice the  velocity in \ref{fig:2layer_2}).
This weak vortical flow, however,  is  unable to fully 
develop into convection before it becomes disrupted by
the interfacial wave below.  Clearly in figure \ref{fig:2layer_3}
the flow in the upper half plane is soon dominated
by the interfacial wave, which by $t=0.042$ has grown quite
large.

\begin{figure}
\epsscale{0.55}
\plotone{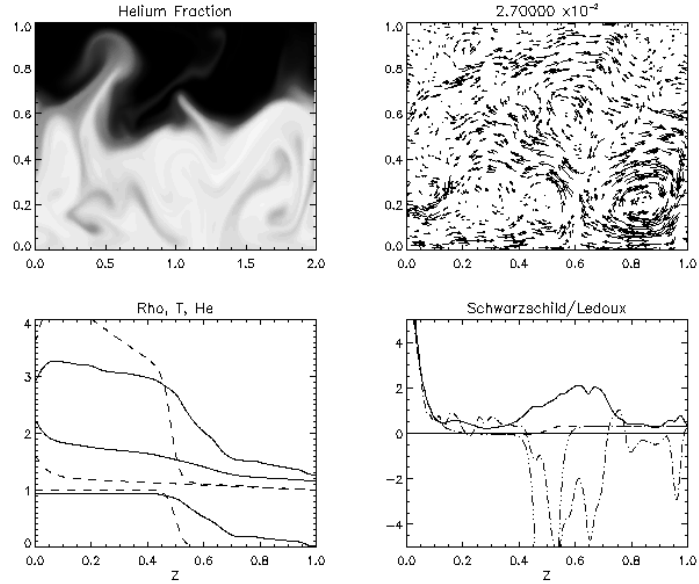}
\caption{A large helium plume results from the self interaction
of relatively small interfacial waves.  The interface is still very
stable.
}
\label{fig:2layer_2}
\end{figure}

Though this simulation  deals with an idealized
setting, it is obvious that the structures in the flow are less simple.  In
particular, the distinction between ``helium mixing plume'', which
results from the interaction of two gravity waves, and a large 
amplitude gravity wave is not precise.  
Gravity waves below an interface are also referred to as
penetrative convection and have been
studied  numerically (see \citet{massaguer},
\citet{hurlburt})
and analytically (\citet{rosner}, 
in the context of a fixed stability interface).
The current results are complicated by the fact that
the stable interface itself interacts with the flow,
yielding a highly nonlinear wave pattern 
and blurring previous distinctions of plume and wave.  
The structure in
figure \ref{fig:2layer_2} is classified as mixing because,
for the most part, it becomes entrained in the uppper fluid.  
The large interfacial wave in figure \ref{fig:2layer_3} is
described as such because, although there is significant
scouring of its upper boundary and entrainment at its crest,
it mostly returns to the lower layer intact.

\begin{figure}
\epsscale{0.5}
\plotone{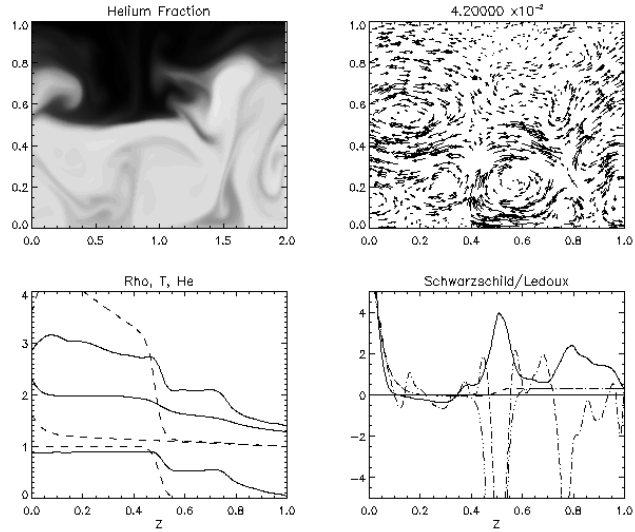}
\caption{Motion in the upper fluid is relatively insignificant
still by $t=0.042$.  A plume driven interfacial wave dominates
the figure and is seen as a second knee ahead of the
interface in the mean field plots.  $\epsilon_L \ll 0$ implying 
that the interface remains stable.
}
\label{fig:2layer_3}
\end{figure}

\begin{figure}
\plotone{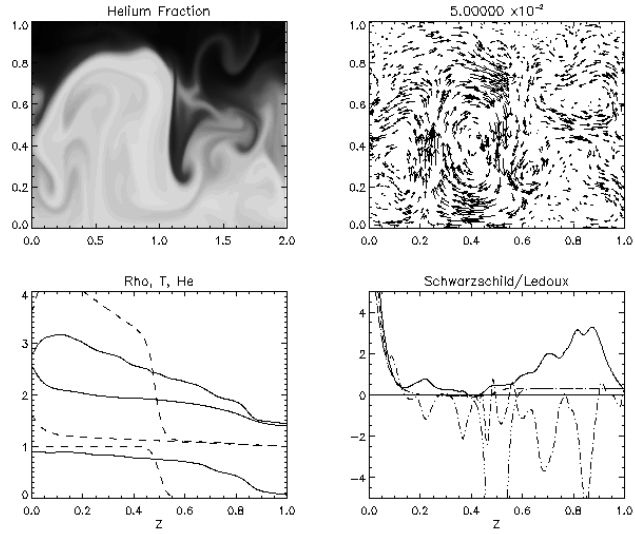}
\caption{The up-flow is dominated by a single
thermal on the bottom boundary which excites a very large
amplitude, large scale interfacial wave.  Though an 
interface is barely apparent in the mean field plotes, it is clear
from the helium concentration that the fluid is not well yet mixed.
}
\label{fig:2layer_4}
\end{figure}

Using the Richardson number as a criterion for 
buoyancy stability, one concludes that 
the interface is stable throughout the simulation
This is in stark contrast to the fact that 
interfacial waves are significant and at $t=0.05$,
figure \ref{fig:2layer_4}, are of amplitude
equal to the layer height.  
As in the prior simulation, this discrepancy results from the
fact that at late times the 
up-flow at the lower boundary  concentrates into one dominant
rising thermal.  
Recall that the Richardson number, plotted in 
figure \ref{fig:2layer_ri} for this case, is the ratio of the buoyancy
across an eddy height to the {\em average} kinetic energy in the
lower layer: a mean field argument.  
Since the buoyancy and resulting momentum of the 
thermal plume is much greater than would be anticipated from a mean field
argument it is able to penetrate much deeper than expected.

From figure \ref{fig:2layer_ri} 
the conclusion should be drawn  that, since the
average kinetic energy in the lower domain is always 
$> 1/40$ the potential energy wall of the interface, 
convection should not penetrate deeply.  However, it is
obvious from figure \ref{fig:2layer_4} that the flow is 
collimated in the strong thermal up-flows, thus the average
kinetic energy is not a good representation of the dynamics
in the lower layer.

Therefore it is also clear from this simulation that
if the lower layer develops a large scale convective 
roll, then the interfacial wave will have
a large scale and correspondingly
large amplitude.  Furthermore, figure 
\ref{fig:2layer_4} dramatically shows that in such instance, 
mixing is dominated by entrainment below the wave trough
and by scouring of the interface from above through the weaker flow
generated by the interfacial wave above it.

\begin{figure}
\epsscale{0.5}
\plotone{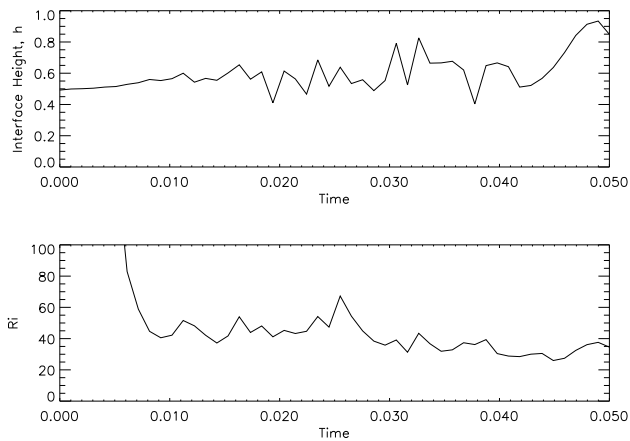}
\caption{The interface height and Richardson number in the
two layer simulation, $VII$.
}
\label{fig:2layer_ri}
\end{figure}

\section{Discussion}

\subsection{Comparison with previous work}
\label{sec:compare}

\begin{figure}
\epsscale{0.55}
\plotone{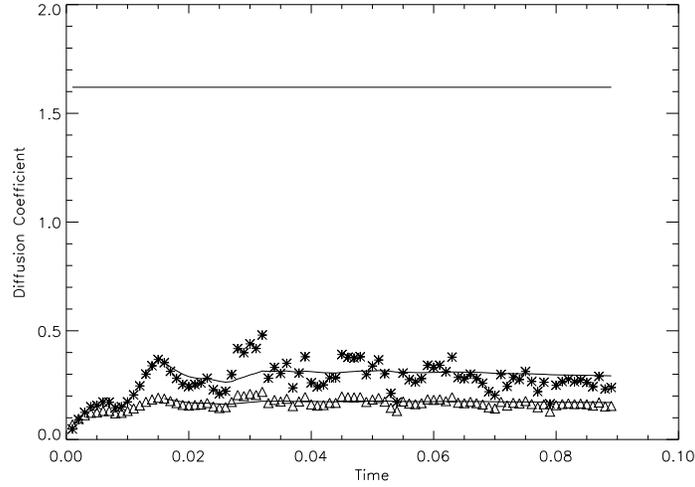}
\caption{Solid line is the calculated diffusion coefficient 
from a $h \propto t^{1/2}$ fit of $V$.  Asterisks are
calculated from \citet{stevenson} while triangles
are calculated from \citet{spruit}; time averages
of the latter two are added.
}
\label{fig:diff_coeff}
\end{figure}

\begin{figure}
\epsscale{0.55}
\plotone{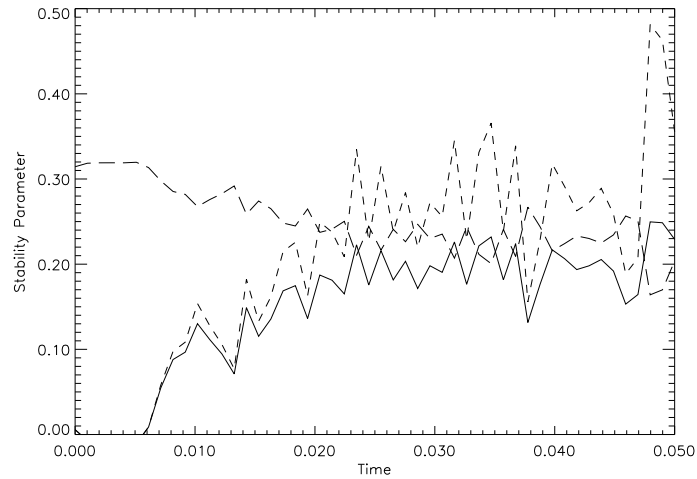}
\caption{The stability parameter for the interface of $VII$
derived using predictions from 
\citet{stevenson} - solid, \citet{fern_89} - dash
and \citet{spruit}- dot-dash.  }
\label{fig:stab_param}
\end{figure}

I remind the reader of the mixing theories of 
\citet{stevenson}, \citet{spruit} and 
\citet{fern_89}, in order to compare them with these simulations.
The former two provide mixing length theories of semiconvection 
in the presence of layers 
and, therefore, effective diffusion coefficients for transport
across these layers.  The latter  derives the
 equilibrium height of the first layer through energy arguments
and provides a criterion for onset of convection above the
first layer.

First let us consider \citet{fern_89}.  
Using conservation of energy and
molecular weight and with the assumption that the kinetic
energy available to convection is proportional to the heat
flux from the boundary, he derived the following
relation
\beq
h_c \approx C
\left[\frac{Ra_{T,\, bot}}{Ra_{He}^{\frac{3}{4}} Pr^{\frac{1}{2}}} \right]^{\frac{1}{2}}
\label{eq:height}
\eeq
for $h_c$, the critical interface height of the first layer.  The constant,
$C$ must be determined experimentally as it contains the parameters
of convective efficiency and
its dependence on $Pr$ is not at all known.
Calibrating the 
numerical experiments to the theory may not be reasonable, but
the  theory can be used to compare the simulation results 
amongst themselves.  
By calibrating the constant to simulation $II$,
where $h_c = 0.8$, equation (\ref{eq:height}) 
yields the critical layer heights
$h_c = 0.50, \,\,1.4,\,\, 0.63,\,\,1.0,\,\, 0.55$ for simulations
$I,\, III (a),\, III (b), \,IV$ and $V$ respectively.  Though not 
quantitatively correct, 
these agree qualitatively with the observation
of a layer in $I$ and $V$, plumes impinging on the upper boundary
in $III (a)$ and interfacial waves hitting the upper boundary in
$IV$. The lack of layering in $III (b)$ cannot, of course,  be explained by
this theory.

Of the necessary criteria suggested at the outset for the
formation of a second layer, only 
simulation $I$ satisfied them both.
Since it would be useful to refine them, 
we must now  turn our consideration to the dynamics
of the interface above the layer.
\citet{fern_89}
reproduced the observation of \citet{turner_68}
that the onset of a second layer
requires  the Rayleigh number in the interfacial region be 
sub-critical to overturning convection.  If the interfacial thickness
is denoted $\delta$ then this amounts to the requirement that 
$\delta \ll h_c$.  Again, using energetic arguments, it can
be shown that
\beq
\frac{\delta}{h_c} = \frac{Ra_{crit}^{\frac{1}{4}}}{\sqrt{2} C}
\left[\frac{ Ra_T Pr}{Ra_{He}} \right]^{\frac{1}{4}}
\label{eq:interface}
\eeq
where the Rayleigh numbers are to be measured at the interface level and
$C$ is the same as in equation (\ref{eq:height}).  The critical
Rayleigh number, $Ra_{crit}$ must also be determined experimentally
and was found to be rather higher than that of classical convection, making
the combination of constants of order one.  

	This can be reworked into necessary condition for the onset
of oscillatory convection above the first layer and, amounts to
\beq
\Sigma_{Fern} \equiv \frac{\epsilon_S}{\nabla_{\mu}} \ll
\frac{ 4 C^4}{Ra_{crit} Pr}.
\label{crit_fern}
\eeq
Again experimental
results suggest that the combination of constants on the right hand side
are order one.

The theories of \citet{spruit} and \citet{stevenson} 
give effective diffusion coefficients for helium transport across
an interface under the respective assumptions of
enhanced diffusion or internal wave breaking.
If the rate of diffusion is greater than the growth
rate of oscillations above the interface, then the region
above will slowly be entrained into the convection below.  Otherwise,
there is some hope for the growth of oscillations above.
The growth rate of weakly forced overstable modes of wavenumber, $k$, is
\beq
\gamma = \frac{\epsilon_S}{2|\epsilon_L|} k^2
\label{eq:growth}
\eeq
while the effective molecular diffusivities are
\beq
\tau_{wave} = \left[\frac{\epsilon_S}{\nabla_{\mu}}\right]^2
\eeq
and
\beq
\tau_{diff} = \sqrt{\tau} \frac{\epsilon_S}{\nabla_{\mu}}
\eeq
for Stevenson's wave model and Spruit's 
enhanced diffusion model, respectively.  The rate of diffusion of
modes $k$ is proportional $k^2 \tau_{eff}$ and the requirement
that oscillations grow more rapidly than a diffusion
time yields the stability criteria
\beq
\Sigma_{wave} \equiv \frac{\epsilon_S}{\nabla_{\mu}} 
\frac{|\epsilon_L|}{\nabla_{\mu}} \ll 1
\label{stab_stev}
\eeq
and
\beq
\Sigma_{diff} \equiv \sqrt{\tau} \frac{|\epsilon_L|}{\nabla_{\mu}} \ll 1
\eeq
respectively for the two theories.  

The effective diffusion coefficients and 
stability parameter predicted by each theory can be 
measured by evaluating $\epsilon_S$ and $\nabla_{\mu}$ at 
the interface.  
Figure \ref{fig:diff_coeff} compares $\tau_{wave}$ and 
$\tau_{diff}$ with the fit $h = 1.8 \sqrt{t} $ to simulation
$V$ which yields $\tau_{eff} \approx 1.62$.  While the
difference of both theories with the numerical data is 
astonishing, the enhanced diffusion model further suffers 
from the fact that it is proportional to $\sqrt{\tau}$, 
a negligibly small number in stars.

The variation of the stability criteria with time 
for simulation $VII$ is plotted in 
figure \ref{fig:stab_param} and they all notably remain
less than one, but not significantly so. 
Again, the criterion $\Sigma_{diff}$ is proportional to $\sqrt{\tau} 
\, (\sim 1/3 $ here)
meaning that, in the limit of small molecular diffusivities, 
layered convection will always occur.  This conclusion simply
results from the assumption that only enhanced diffusive transport
acts across the interfaces, inconsistent with the wave breaking seen in 
the simulations.

Both $\Sigma_{Fern}$ and $\Sigma_{wave}$
were derived considering buoyancy effects, yet the fact 
that they  agree well throughout
most of the  simulation is still striking.
This agreement results from
the fact that the Ledoux criterion is dominated by the 
compositional gradient at the interface, $|\epsilon_L| \sim \nabla_{\mu}$.  
Nonetheless, it is clear that $\tau_{wave}$ underestimates the
true layer entrainment and thus the stability criterion is overestimating
the stability of the layer.  Therefore, it seems unlikely that a second
convecting region will form above the interface in $VII$ even though
the Richardson number criteria indicates that it should be stable.

\subsection{Conclusions and future work}

These  simulations have reproduced layer formation  in the context
of laboratory flows and distinguish the mechanisms for 
transport across a compositional interface:  enhanced diffusion,
scouring and wave breaking.  
They also confirm  the difficulty of forming layers
in a low Prandtl number fluid.  Characteristic of all 
of the results is the inverse cascade of velocity to large length
scales resulting in strong isolated upward rising thermals 
which are able to excite 
high amplitude, long wavelength waves on the interface.  Mixing across the
interface is dominated by wave breaking in all of the low Prandtl
number cases; even, and most importantly, when the 
interface is prescribed and highly stable ($VII$).  

Though it remains to be determined whether strong, isolated
thermals develop in a three dimensional fluid, it can yet be
safely concluded that, at low Prandtl number, cross interfacial
mixing will be dominated by wave breaking.  This is especially
evident at early times in $VII$ where even small interfacial 
waves drive large ``splashes'', mixing helium high into the stable
layer.  Therefore, the criteria 
associated with the \citet{stevenson} or 
\citet{fern_89} 
theories of interfacial mixing seem to be the most appropriate
for determining whether a second layer, and thus
multiple layers, will form.  Even so, these theories greatly
underestimate the mixing across the interface.

This underestimate is linked to the the large interfacial
waves and therefore the critical Richardson number for the
stability of an interface.  Comparing simulation $V$ 
to the low Prandtl number case, $I$,  it is clear
that the critical Richardson number must 
increase with decreasing Prandtl number.  

Further study of whether the estimate of \citet{stevenson}
truly underestimates mixing requires determining whether or
not the large interfacial waves persist in 3-dimensional simulations
and at large aspect ratio.
Though the former would  be most valuable, a parameter space study is 
not computationally feasible.  The best prospects for sorting 
out the issue of isolated thermals rests in  continuing 
this work at larger aspect ratios.

The simulations I have presented envision  a  scenario where an
energy flux is abruptly turned on below a region
which is marginally unstable to semiconvection.  This results
in the formation of layered convection due
to the redistribution of a compositional gradient 
in a thin layer and on time scales short 
compared to a thermal diffusion time.  
The convection within such a layer is concentrated in plumes
spaced such that the largest rolls have aspect ratio 
approximately $1$.  The interface itself is characterized
by large internal waves and smaller breaking waves, 
making it ineffective as a barrier to transport.  
Since, in all cases considered, the fluid is very well
mixed after a small fraction of a thermal diffusion time,
it is unlikely that layering caused by buoyancy disruption
would play any role in stellar convection.  In fact, the 
very large effective diffusion measured herein suggests that such
layered structures are so unstable  as to make them dynamically
unimportant even when 
a compositional gradient is redistributed by other mechanisms
such as rotation or large internal waves, which set up initial
conditions similar to $VII$.

\acknowledgements

I gratefully acknowledge support and computing resources
from the ASCI/Alliances Center for Astrophysical
Thermonuclear Flashes at the University of Chicago under DOE subcontract
B341495.
Preliminary analytic studies were also 
supported by an NSF Graduate Fellowship
and NASA Space Theory Grant NAG5-8495.  
This work was motivated by suggestions
of Prof. Robert Rosner and benefited greatly from continuing conversations
with him.  Without Dr. Francesco Rubini, who provided the numerical 
expertise and code, this project would have taken an extra year.  Without
Dr. Diana Hall, this project would have taken six months less.


\end{document}